\documentclass[sigconf]{acmart}

\copyrightyear{2024}
\acmYear{2024}
\setcopyright{acmlicensed}\acmConference[SIGGRAPH Conference Papers '24]{Special Interest Group on Computer Graphics and Interactive Techniques Conference Conference Papers '24}{July 27-August 1, 2024}{Denver, CO, USA}
\acmBooktitle{Special Interest Group on Computer Graphics and Interactive Techniques Conference Conference Papers '24 (SIGGRAPH Conference Papers '24), July 27-August 1, 2024, Denver, CO, USA}
\acmDOI{10.1145/3641519.3657440}
\acmISBN{979-8-4007-0525-0/24/07}

\usepackage{graphicx}
\usepackage{amsfonts}
\usepackage{amsmath}
\usepackage{booktabs}
\usepackage{xcolor}
\usepackage{multirow}
\usepackage[normalem]{ulem}
\usepackage[super]{nth}
\usepackage{cleveref}
\usepackage{colortbl}
\usepackage{xspace}

\usepackage{wrapfig}

\newcommand{\todo}[1]{{\color{red}#1}}

\newcommand{\revise}[1]{{\color{black}#1}}

\newcommand{\quotes}[1]{``#1''}

\acmSubmissionID{588}

\newcommand{\dataset}{ \Omega}

\newcommand{\nbframepred}{ F }
\newcommand{\nbframepast}{ P }
\newcommand{\nbapplied}{ M }

\newcommand{\control}{ \mathbf{c}}

\newcommand{\rootglobal}{ \textbf{O} }
\newcommand{\rootlocal}{ \textbf{V} }

\newcommand{\sample}{ \mathbf{x} }
\newcommand{\samplepred}{ \hat{\mathbf{x}} }
\newcommand{\net}{ \mathcal{G} }
\newcommand{\nbdiffsteps}{ T }

\newcommand{\pastmotion}{ \mathbf{p} }
\newcommand{\stylelabel}{ \control_l}
\newcommand{\rootposproj}{ \control_{rv} }
\newcommand{\rootrotproj}{ \control_{ro} }

\newcommand{\skeleton}{ \mathbf{S} }
\newcommand{\loss}{ \mathcal{L} }
\newcommand{\losssamp}{ \loss_{\mathrm{samp.}} }
\newcommand{\losspos}{ \loss_{\mathrm{pos.}} }
\newcommand{\lossfoot}{ \loss_{\mathrm{foot}} }
\newcommand{\lossvel}{ \loss_{\mathrm{vel.}} }
\newcommand{\lambdasamp}{ \lambda_{\mathrm{samp.}} }
\newcommand{\lambdapos}{ \lambda_{\mathrm{pos.}} }
\newcommand{\lambdafoot}{ \lambda_{\mathrm{foot}} }
\newcommand{\lambdavel}{ \lambda_{\mathrm{vel.}} }

\newcommand{\nbextension}{ K }

\newcommand{\nbjoint}{ J }
\newcommand{\rotdim}{ Q }
\newcommand{\realnum}{ \mathbb{R} }

\newcommand{\jointrot}{ \textbf{R} }

\setcitestyle{nosort,square}

\begin{document}

\title{Taming Diffusion Probabilistic Models for Character Control}

\newcommand{\name}{CAMDM\xspace}

\author{Rui Chen}
\email{riorui@foxmail.com}
\orcid{0009-0003-7122-5207}
\affiliation{%
  \institution{Hong Kong University of Science and Technology}
  \country{Hong Kong}
}
\authornote{Joint first authors and work done during an internship at Tencent AI Lab}

\author{Mingyi Shi}
\email{myshi@cs.hku.hk}
\orcid{0000-0002-5180-600X}
\affiliation{%
  \institution{The University of Hong Kong}
  \country{Hong Kong}
}
\authornotemark[1]

\author{Shaoli Huang}
\email{shaolihuang@tencent.com}
\orcid{0000-0002-1445-3196}
\affiliation{%
  \institution{Tencent AI Lab}
  \country{China}
}

\author{Ping Tan}
\email{pingtan@ust.hk}
\orcid{0000-0002-4506-6973}
\affiliation{%
  \institution{Hong Kong University of Science and Technology}
  \country{Hong Kong}
}

\author{Taku Komura}
\email{taku@cs.hku.hk}
\orcid{0000-0002-2729-5860}
\affiliation{%
  \institution{The University of Hong Kong}
  \country{Hong Kong}
}

\author{Xuelin Chen}
\email{xuelin.chen.3d@gmail.com}
\orcid{0009-0007-0158-9469}
\affiliation{%
  \institution{Tencent AI Lab}
  \country{China}
}
\authornote{Corresponding author}

\renewcommand{\shortauthors}{\name}

\definecolor{red}{rgb}{0.8, 0.2, 0.2}
\definecolor{purple}{rgb}{0.99,0.2,0.72}

\newif\ifdraft
\drafttrue

\ifdraft
\newcommand{\xl}[1]{{\color{blue}[xl: #1]}}

\else
\newcommand{\xl}[1]{}
\newcommand{\todo}[1]{}

\fi
\begin{abstract}

We present a novel character control framework that effectively utilizes motion diffusion probabilistic models to generate high-quality and diverse character animations, responding in real-time to a variety of dynamic user-supplied control signals. 
At the heart of our method lies a transformer-based Conditional Autoregressive Motion Diffusion Model (\name), which takes as input the character’s historical motion and can generate a range of diverse potential future motions conditioned on high-level, coarse user control.
To meet the demands for diversity, controllability, and computational efficiency required by a real-time controller,
we incorporate several key algorithmic designs. 
These include separate condition tokenization, classifier-free guidance on past motion, and heuristic future trajectory extension, all designed to address the challenges associated with taming motion diffusion probabilistic models for character control.
As a result,
our work represents the first model
that enables real-time generation of high-quality, diverse character animations based on user interactive control, supporting animating the character in multiple styles with a single
unified model. 
We evaluate our method on a diverse set of locomotion skills, demonstrating the merits of our method over existing character controllers.

\end{abstract}



\begin{teaserfigure}
\center
  \includegraphics[width=\textwidth]{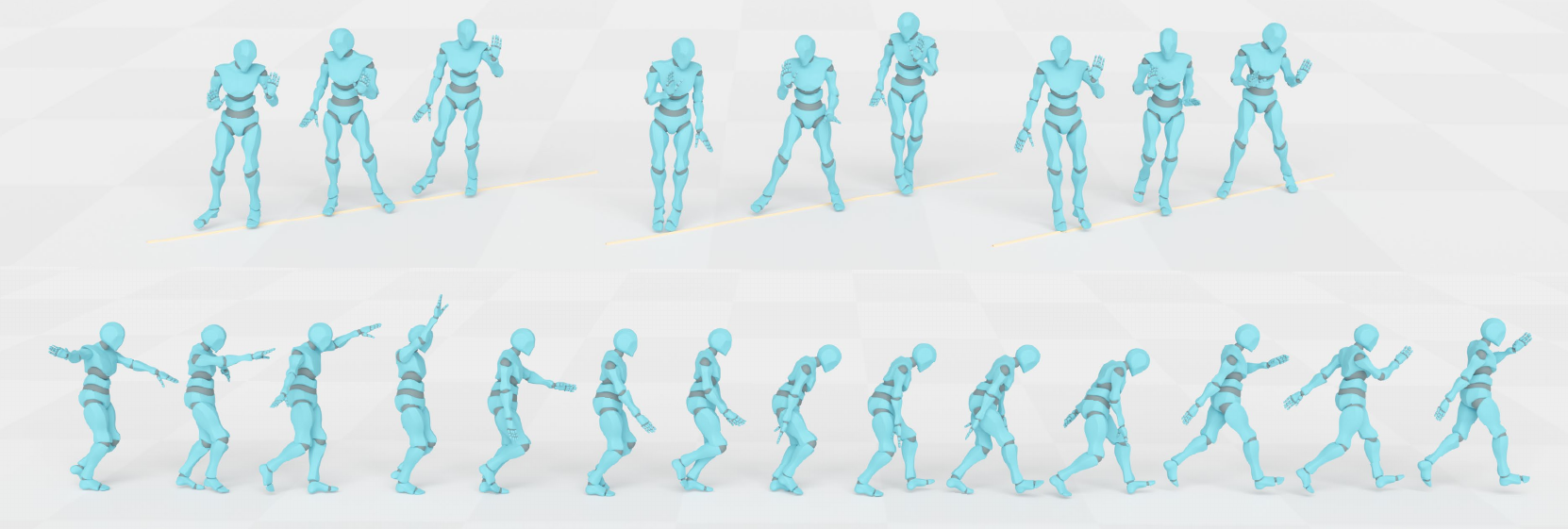}
  \caption{
  Real-time character control powered by our Conditional Autoregressive Motion Diffusion Model (\name). 
  Top: Our method produces diverse outputs for character animation even given identical input control signals.
  Bottom: Our method can generate seamless transitions between different styles, even in cases where the transition data is absent from the dataset.
  All achieved with a single unified model, in real time.
  }
  \Description{
  TBA.
  }
  \label{fig:teaser}
\end{teaserfigure}

\setcopyright{acmlicensed}
\acmJournal{TOG}
\acmYear{2023} \acmVolume{42} \acmNumber{4} \acmArticle{} \acmMonth{8} \acmPrice{15.00}\acmDOI{10.1145/3592395}

\maketitle

\section{Introduction}

Real-time character control is the holy grail in computer animation, playing a pivotal role in creating immersive and interactive experiences. 
It involves generating natural and expressive motions of virtual characters, that align well to the high-level specifications of the user interactions in a dynamic and real-time environment.
The evolution of this field has been driven by remarkable advancements in data-driven techniques~\cite{motionfield, mm} and the growing demand for realistic and interactive content in various sectors, including gaming and virtual reality.

%
Despite significant progress, real-time character control continues to present several key challenges.
These include
generating \emph{visually realistic} motions of high quality and diversity, 
ensuring \emph{controllability} of the generation, and 
attaining a delicate balance between \emph{computational efficiency} and visual realism.
More specifically,
\emph{diversity} can refer to variations of different granularity, including intra-style diversity, which requires modeling the variation of joint curves within a particular style, and inter-style diversity, which refers to the variation between different styles. 
While much progress has been made in this field, almost all existing approaches struggle to address one or more of these challenges, consequently impeding their widespread adoption and implementation.

Recent advancements in deep learning, particularly diffusion-based generative models, have greatly impacted the motion generation field, showing more promise in potentially resolving these challenges.
As one of the pioneering works in this direction,
Phase-Functioned Neural Network (PFNN)~\cite{pfnn}, along with the following studies~\cite{mann, deepphase, localphase}, learns a deep neural network to predict the next posture of the character given user control signals, in an autoregressive manner.
However, 
defining the future posture can be vague based on high-level, \quotes{sketchy} user inputs.
Therefore, the deterministic models employed in their approach suffer from regression to the mean pose, resulting in artifacts (e.g., sliding feet and reduced range of motions), despite the inclusion of heuristically designed or learned features for disambiguation. 
Moreover, the lack of motion diversity leads to repetitive and visually monotonous characters, reducing the visual realism in real-time and interactive applications.
More recently,
motion diffusion probabilistic models have shown the capability of capturing complex data distribution in large motion datasets and then generating high-quality and diverse samples in highly controllable ways through a long Markov denoising chain~\revise{~\cite{mdm, zhang2022motiondiffuse, jiang2023motiongpt, alexanderson2023listen, Wang_2023_ICCV, chen2023executing, priormdm, omnicontrol, gmd}}.
Now, the question arises as to how diffusion-based generative models can be effectively employed in real-time character controllers, producing high-quality and diverse outputs in response to user interactions.

In this paper, 
we introduce a novel character controller, which effectively harnesses motion diffusion probabilistic models for producing character animations that appear natural and respond in real-time to various, varying user-supplied control signals.
These control parameters specify various aspects to be expected from the generated motions, including the style/gait, moving speed, facing direction, future trajectory on the ground, etc.
Our method builds upon the autoregressive framework introduced in~\cite{pfnn},
but the core of our method is a transformer-based \text{Conditional Autoregressive Motion Diffusion Model} (CAMDM, pronounced \quotes{cam-dem}), 
which takes as input the character's history motions and can generate a range of diverse potential future motions conditioned on high-level, coarse control signals.
Particularly,
to achieve the controllability, diversity, and computational efficiency for real-time character controllers as aforementioned,
several algorithmic designs are incorporated to address challenges that arise from working with motion diffusion models.

First,
in contrast to previous approaches that represent various conditions as a single feature vector~\cite{pfnn, mdm},
which can lead to unstable control due to certain feature components dominating unexpectedly,
our model learns a separate token for each condition. 
By leveraging the attention mechanism of the transformer, our model enhances the effectiveness of each condition within the network, resulting in stable control.
Second,
regarding the diversity,
on the one hand,
we found training the model to forecast long future motions and applying as many frames as possible to the character (only when the user inputs remain unchanged) significantly improved the intra-style diversity of the autoregression.
Nonetheless, during runtime this strategy causes missing model-predicted future trajectories, which are supposed to be blended with user-supplied synthetic future trajectories for balancing the motion naturalness and trajectory alignment of the next prediction~\cite{pfnn},
leading to abrupt jittering between the past and predicted motions.
We then devise a simple yet effective module, which recycles the last small fraction of the previous model-predicted future trajectory to obtain a heuristic extension beyond the end of the model-predicted future
trajectory. 
On the other hand,
transitions between different styles are generally lacking in the mocap data, which poses challenges to controlling the style and thus obtaining inter-style diversity in the autoregression.
Hence, we propose to train the autoregressive motion diffusion with classifier-free guidance~\cite{cfg} on the conditional token of history motions, instead of on the token of style labels \--- an intuitive choice but fails in our experiments.
We show that this design significantly helps transition between different styles in the autoregression, yielding natural transitional motions.
Last,
our model uses only 8 denoising diffusion steps, which is significantly less than the 1000 steps in other literature, and hence can respond to the control in real-time.

To the best of our knowledge, our work represents the first model that enables real-time generation of high-quality and diverse character animations based on user interactive control, supporting animating the character in multiple styles with a single unified model.
%
We demonstrate our method on a large publicly available mocap dataset featuring a diverse set of locomotion skills, as locomotion skills are important ingredients of character animations in many applications that involve user interactive controls.
Our extensive experimental results show
that our method is more than competent in producing high-quality and diverse character animations in real-time response to user-supplied control signals. 
Important algorithmic designs are also validated in ablation studies.

\section{Related work}
In this section,
we first review deep learning-based approaches for real-time locomotion generation based on user-supplied control parameters. 
Next, we review recent literature for synthesizing natural and expressive motions using diffusion probabilistic models.

\paragraph{DL-based Real-time Character Controllers}
Real-time character control plays an important role in character animation in gaming.
Generally, state machines and motion matching are commonly used techniques in the industry.
The former defines a set of states characters can be in, such as idle, walking, running, jumping, attacking, etc. Each state represents a specific animation and behavior. 
Transitions between states are triggered by certain user inputs. 
Motion matching, on the other hand, searches and blends motion data in a large dataset to create smooth and realistic character animations based on local control context.
While these techniques have allowed for managing character behaviors and interactions,
they face challenges in delivering high-quality and diverse animations, and in scaling up to large-scale mocap data for producing much more complex character animations in controllable ways.
Hence, research efforts have taken various approaches to apply deep neural networks to large-scale mocap data, due to their scalability and high run-time performance~\cite{learnedmm, motionrecom}.

Training neural networks to predict future postures of the character based on user inputs is still challenging,
as the high-level, \quotes{sketchy} user inputs do not encompass every fine detail, implying ambiguities in the final definition of the expected motion.
Hence, simply training the network with the MSE minimization objective would result in obvious artifacts in the character animation, e.g., foot skating.  %
Some propose to overcome this issue with the help of additional features crafted from the motion data.
\citet{pfnn}
propose the phase-functioned neural network, which learns and predicts the phase of the character's feet, significantly disambiguating the future posture prediction in locomotion.
\citet{localphase} extend the concept of motion phase to complex motion by assigning each bone motion in a character with a separate motion phase. 
\citet{mann} propose a method that utilizes autoregressive and control information to modify network weights, enabling the generation of controllable quadruped motion. They demonstrate the successful generation of both cyclic motion, such as gait, and simple non-cyclic motion like jumping.
\citet{deepphase} extract a deep multi-dimensional phase space from full-body motion data to achieve better temporal and spatial alignment and demonstrate the merit of the learned deep features in character animation.
However, 
their character control framework is based on deterministic models, which inherently face challenges in generating realistic motions, especially when aiming for high diversity. This struggle becomes more pronounced when the available predictive input information is inadequate for effective disambiguation.

Probabilistic models offer an alternative to prevent collapsing into a mean pose. 
By training models to learn all plausible poses based on history motions and control inputs, the generated motion should exhibit convincing and realistic motion while following the user's specifications.
\citet{motionvae} first train a Variational Auto-Encoder (VAE) to learn a latent motion space,
with which a reinforcement learning policy network learns to steer the latent variable to follow control inputs.
However, their method often encounters significant errors in accurately tracking input trajectories, primarily due to the limitations of VAEs in effectively modeling complex data distributions.  
While the probabilistic normalizing flow-based method introduced in \cite{moglow} can model conditional motion distributions and thus accommodate user-supplied control parameters,  its real-time performance is limited, implying it is less suitable for interactive applications like video games.

\revise{
Another line of literature focuses on producing physically plausible motions,
where typically deep reinforcement learning (deep RL) policy networks are trained to control characters in simulated environments.
\citet{peng2018deepmimic} demonstrate the potential of deep RL in animating characters within physically simulated environments by training an RL policy to adhere to a reference kinematic motion.
Subsequent works propose controlling physically simulated characters via RL policies trained to track a wide range of reference motions produced by kinematic motion generation modules~\cite{bergamin2019drecon, won2020scalable, fussell2021supertrack}.
However, these methods typically rely heavily on the kinematic motion generation module, leading to challenges in composing diverse motions that surpass the capabilities of the kinematic generation.
%
Moreover, probabilistic models have been leveraged to enhance the diversity and generalizability of physically simulated characters~\cite{won2022physics, peng2021amp, peng2022ase, yao2022controlvae}, enabling diverse generation of physically plausible motions given only a limited set of kinematic motions.
Although some methods in this domain have also been introduced to learn character controllers from extensive datasets~\cite{tessler2023calm, dou2023c}, 
these techniques still face difficulties in meeting all the essential criteria of a real-time character controller, namely realism, controllability, and computational efficiency as mentioned above.

}

\paragraph{Motion Diffusion Probabilistic Models}
Recently, diffusion-based generative models have revolutionized natural image generation by learning to denoise~\citep{glide, zero, saharia2022photorealistic, rombach2022high}. As a result, there has been a surge of subsequent works adopting these models for motion generation~\cite{mdm, zhang2022motiondiffuse, kim2023flame, dabral2023mofusion, jiang2023motiongpt, alexanderson2023listen, yuan2023physdiff, Wang_2023_ICCV, chen2023executing}, leading to state-of-the-art performance in generating diverse and natural human motions conditioned on arbitrary textual prompts.

Although significant progress has been made in these endeavors, the language description alone has inherent limitations in providing more precise control over the generated motion. 
Hence, a new line of research has emerged, focusing on incorporating user-provided spatial constraints to enhance control over motion generation~\cite{gmd, priormdm, omnicontrol}.
Our work also builds upon the powerful diffusion probabilistic model for conditional motion generation,
but differs significantly from these existing models by the particular emphasis on learning a conditional autoregressive motion diffusion model for real-time character control.
Another related work~\cite{amdm} has showcased the utilization of an autoregressive motion diffusion model to generate future character postures based on joystick control signals. However, they have not demonstrated the ability to support multiple styles and their approach achieves controllability by training a dedicated policy network through reinforcement learning to guide the motion diffusion synthesis process.
In contrast,
our method directly models conditional motion distributions, and so has controllability built in.
\section{Method}
Given a large-scale locomotion dataset $\dataset$,
our method first trains a Conditional Autoregressive Motion Diffusion Model (CAMDM), which
takes as input the past motion $\pastmotion$ of $\nbframepast$ frames of the character, and user control parameters $\control$, and then learns to capture the conditional distribution of the future motions $\sample$ (of $\nbframepred$ frames).
During runtime, \name is applied at each frame with the on-the-fly collected character's historical poses, user control inputs, and randomly sampled Gaussian noise, and then sample from the conditional motion distribution, obtaining a sequence of realistic future postures to be displayed.
 %
The character animated using this autoregressive generation approach can exhibit coherent and diverse motions while adhering to user inputs. 
This is achieved because \name is trained to capture all possible future motions under different conditions, yielding a wide range of plausible animations.

\subsection{Motion Diffusion Model}

Following the recent success of motion generative models, 
our method employs the diffusion probabilistic model, which learns 
to generate by learning to denoise, to model the complex motion data distribution in a large-scale mocap dataset.
Given a clean motion sequence sample $\sample_0$, the forward diffusion process is a Markov chain that progressively adds noise to $\sample_0$ in $T$ steps, producing an array of noisy samples $\sample_1, ..., \sample_T$: $q_\theta(\sample_t|\sample_0) = \mathcal{N}(\sample_t; \sqrt{\bar{\alpha}_t}\sample_0,(1 - \bar{\alpha}) \mathbf{I})$, where $\bar{\alpha}_t = \prod_{s=1}^t(1-\beta_s)$ and $\beta_t$ is the variance scheduler of the added noise.
Then a model parameterized by a deep neural network $\net$ is trained to learn the reverse process in another Markov chain, i.e., learning the mapping $p(\sample_{t-1}|\sample_t)$, to denoise the noisy sample sequentially in $\nbdiffsteps$ steps. 
%
Instead of predicting the added noise as in~\cite{ddpm_ho}, 
we directly predict the clean sample itself at each denoising step: $\samplepred_0 = \net(\sample_t, t)$,
which is a common practice in various diffusion models~\cite{ramesh2022hierarchical, mdm, omnicontrol} and has been shown to
achieve slightly better generative performance by providing additional geometric losses on the predicted results.
Once trained,
$\net$ can transform randomly sampled Gaussian noise sequences to generate high-quality and diverse motion samples through $\nbdiffsteps$ denoising steps.

\begin{figure}[t!]
  \centering
  \includegraphics[width=\linewidth]{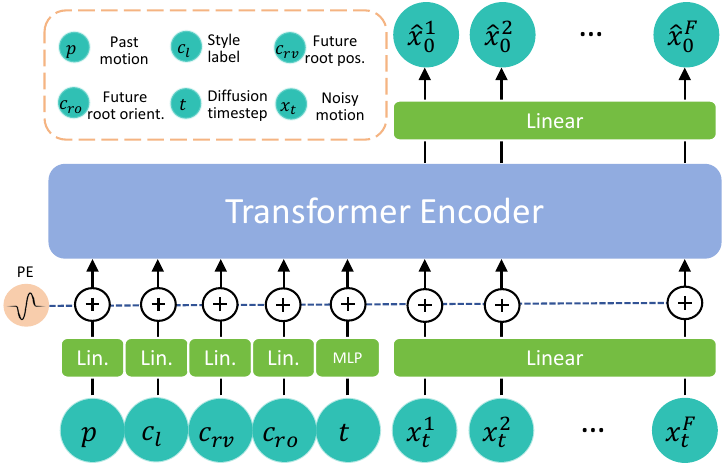}
  \caption{
  Conditional Autoregressive Motion Diffusion Model (CAMDM).
  At each denoising step, the model takes as input a noisy motion sample $\sample_t$, diffusion step $t$, along with various conditions including the past motion $\pastmotion$, style label $\stylelabel$, future root displacements $\rootposproj$, and orientations $\rootrotproj$ projected onto the ground,
  and learns to predict the original clean $\samplepred_0$.
  }
  \label{fig:pipeline}
\end{figure}

\paragraph{Motion Representation}
A motion sample $\sample$ to be predicted by the model is defined by a temporal set of $\nbframepred$ poses
that each $\sample^{1:F}$ consists of root joint displacements $\rootglobal \in \realnum^{\nbframepred \times 3}$ and joint rotations $\jointrot \in \realnum^{\nbframepred \times \nbjoint \rotdim}$,
where $\nbjoint$ is the number of joints and $\rotdim$ is the number of rotation features.
Instead of directly using the global root displacements $\rootglobal$, 
we convert $\rootglobal$ to local root displacements $\rootlocal$
by subtracting the displacement of the first pose from the sequential displacements.
The joint rotations are defined in the coordinate frame of their parent in the kinematic chain,
and we use the 6D rotation representation (i.e., $\rotdim = 6$) proposed by \citet{zhou6d}.

\begin{figure}[t!]
  \centering
  \includegraphics[width=\linewidth]{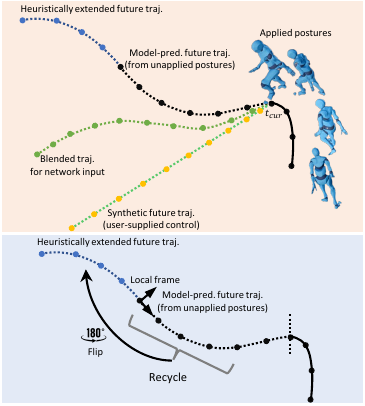}
  \caption{
  Illustration of heuristic future trajectory extension.
  \revise{Top:
  At $t_{cur}$, when the user input changes and autoregression is triggered, the model-predicted future trajectory is shorter than the user-supplied synthetic future trajectory as multiple postures in the generation have been applied to the character. This could result in a blending of two trajectories of different lengths, consequently causing sudden jittering in the next generation if not properly dressed.
  Bottom:
  Assuming that the model generates $\nbframepred = 10$ frames, we simply reuse the last $\nbextension = 4$ of the predicted future trajectory multiple times, if necessary, until the length matches that of the user-supplied synthetic future trajectory.
  For each recycle, we establish a local frame at the last trajectory point based on its position and orientation, then flip the position and copy the orientation of the last $\nbextension$ points to extend the trajectory.
  }
  }
  \label{fig:trajectory}
\end{figure}

\subsection{Conditional Autoregressive Motion Diffusion Model}
Our central objective is to learn a motion diffusion probabilistic model that can take as input the historical poses of the character and output a diverse set of plausible future poses based on user inputs,
so during runtime, the character can be naturally animated to follow control specifications in an autoregressive manner.
To this end,
we further condition the motion diffusion model on additional inputs, including the character's past motion $\pastmotion$ and user-supplied control parameters $\control$, and train it to generate plausible future motions $\sample$ of $\nbframepred$ frames.
More specifically,
the control parameters supported in our method contain the style/gait label $\stylelabel$, future root displacements $\rootposproj$, and orientations $\rootrotproj$ projected onto the ground.
At each denoising step, the model takes as input a noisy motion sample $\sample_t$, diffusion step $t$, along with various conditions,
then learns to predict the original clean $\samplepred_0$ (See Figure~\ref{fig:pipeline}):

\begin{equation}
\begin{aligned}
	\samplepred_0 = \net(\sample_t, t; \pastmotion, \stylelabel, \rootposproj, \rootrotproj).
\end{aligned}
\end{equation}

\paragraph{Training Losses}
The denoising objective is to minimize the MSE loss between the predicted $\hat{x}_0$ and the clean sample $x_0$:

\begin{equation}
	\losssamp = \mathbb{E}
    ||\hat{\sample}_0 - \sample_0 
 ||_2^2.
\end{equation}  
As mentioned in~\cite{mdm}, an advantage of predicting the signal itself is that we can leverage geometric losses. 
During training, we employ differentiable forward kinematics functions $\mathrm{FK}(\cdot)$ to transform local joint rotations into global joint positions,
based on which we can calculate 3D joint position loss:
\begin{equation}
	\losspos=
 \left\|
 \mathrm{FK}(
    \hat{\sample}_0,
    \skeleton)-
    \mathrm{FK}(
    \sample_0,
    \skeleton)\right\|_2^2,
\end{equation}
where $\skeleton$ is the rest skeleton used in the mocap data and remains unchanged throughout the training. 
Similarly, we can also compute the foot contact loss $\lossfoot$ and velocity loss $\lossvel$ during the training (see ~\cite{mdm} for more details).
Finally, the total training loss is written as:
\begin{equation}
\loss = 
\lambdasamp \losssamp + \lambdapos \losspos +
\lambdafoot \lossfoot + 
\lambdavel \lossvel,
\end{equation} 
with $\lambda$ weights controlling the importance of each loss.
Once trained, this conditional autoregressive model can be used in the runtime animation system for character control.

Nonetheless, 
the above description only provides a conceptually valid and general formulation for learning a conditional motion distribution. 
We further elaborate on
algorithmic designs that are crucial for \emph{taming} the motion diffusion probabilistic model for real-time character control.

\paragraph{Separate Condition Tokenization (SCT)}
We found that learning to map various conditions into a single feature vector as in~\cite{pfnn} or a single token as in ~\cite{mdm} can lead to highly unstable control due to certain feature components unexpectedly dominating the regression process. 
For instance, the style label may fail to effectively control the gait of the motion, or the character may not correctly follow the input future trajectory.
Hence,  we choose to use separate tokenizers to learn individual tokens for each input condition and add all these conditional tokens at the beginning of the noisy motion sequence as input to the denoising transformer network.
Then by leveraging the attention mechanism of the transformer, our model enhances the effectiveness of each condition within the network, resulting in a stable control.

\paragraph{Classifier-free Guidance on Past Motion \revise{(CFG-PM)}}
As mentioned earlier, generating transitional motions between different styles is challenging due to the lack of transitional motions in the mocap dataset.
While one workaround is to utilize the classifier-free guidance (CFG) technique with style label conditions~\cite{alexanderson2023listen}, this strategy fails to effectively transition between different styles in our tasks. 
We discovered that the key lies in controlling the influence of the \emph{past motion} in conditional autoregressive generation, which led us to apply CFG to the conditional past motion. 
Specifically, during training, we randomly set the conditional past motion token to nil with a probability of 0.15.
At runtime,
when the style label is changed to a new one by the user input,
the future motion is sampled with a guidance scale of $\gamma$:
\begin{equation}
\begin{aligned}
	\net(\sample_t, t; \pastmotion, \stylelabel, \rootposproj, \rootrotproj) = 
    \net(\sample_t, t; \pastmotion=\varnothing, \stylelabel, \rootposproj, \rootrotproj) \\
    +~\gamma \bigl( 
    \net(\sample_t, t; \pastmotion, \stylelabel, \rootposproj, \rootrotproj) - \net(\sample_t, t; \pastmotion=\varnothing, \stylelabel, \rootposproj, \rootrotproj) 
    \bigr).
\end{aligned}
\end{equation}
\revise{Commonly, CFG is typically utilized with $\gamma>1$ to amplify the conditioning. However, in this paper, we employ $\gamma<1$ to mitigate the influence of past motion while achieving smooth transitions between different styles. 
Note that CFG-PM is NOT used when the style label remains unchanged. Refer to section \ref{sec:imp} for more details.}

\paragraph{Heuristic Future Trajectory Extension (HFTE)}
Regarding the intra-style motion diversity mentioned earlier,
%
we found the \quotes{lazy} trigger of the next autoregression significantly improved the intra-style diversity of the autoregression.
The \quotes{lazy} trigger means that we train the model to forecast a long future motion, and then during runtime apply as many frames as possible to the character when the user-supplied control remains unchanged.
%
%
%
\revise{Specifically, our method is based on the character control framework in~\cite{pfnn}, where the \emph{model-predicted future trajectory} (in our case the floor projection of the root joint of generated motions) is blended with the user-supplied \emph{synthetic future trajectory},
which is synthesized by the critical spring damper function~\cite{mm},
for obtaining the final controlling trajectory input to the network. This strategy offers a good balance between the motion smoothness and trajectory alignment of the next prediction.
Previous methods~\cite{pfnn, mm} adopting this strategy usually generate and apply only one posture to the character at a time, whereas we generate and apply $\nbapplied$ postures from the generated motion before the next autoregression.
Obviously,
this \quotes{lazy} trigger strategy can deplete the model-predicted future trajectory.
This can result in a blending of two trajectories of different lengths, consequently causing sudden jittering between the current posture and the next autoregressed posture if not properly dressed (see the top of Figure~\ref{fig:trajectory}.
%
%
We then devise a simple yet effective module, which
heuristically constructs a smoothly extended trajectory beyond the end of the model-predicted future trajectory,
by
recycling the last $\nbextension$ points of the model-predicted future trajectory backward and forward multiple times, if necessary, until the length matches that of the user-supplied synthetic future trajectory. See the bottom of Figure~\ref{fig:trajectory}.
}
%

\paragraph{Real-time Generation.}
Our task requires the model to respond to the user control in real time.
Although there is extensive research in speeding up the diffusion model~\revise{~\cite{ddim, cm, lcm, meng2023distillation}}, we observed that their performance on our conditional autoregressive motion diffusion resulted in degraded generation results.
Similar to the finding in~\cite{amdm} we empirically discovered that using only 8 diffusion steps during both training and runtime surprisingly leads to convincing performance.

\subsection{Implementation Details}
\label{sec:imp}

\paragraph{Training.}
The mocap data utilized for training our model is at a frame rate of 30 frames per second.
By default, our model is conditioned on $\nbframepast = 10$ frames of the past motion and learns to generate $\nbframepred = 45$ frames for the future motion.
We set $\lambdasamp = \lambdapos = \lambdafoot = \lambdavel = 1$ when training the denoising model.
We implement the denoising network with a straightforward transformer encoder-only architecture and follow the hyperparameters described in~\cite{mdm} for training. Our model is trained on a single A100 GPU for about 20 hours, and the character controller animates the character in real time on a RTX 3060 GPU.

\paragraph{Runtime.}
Our runtime system is based on the open-source system AI4Animation~\cite{ai4animation}, and we deploy the trained model using the third-party library ONNX~\cite{onnx} for real-time inference.
The file containing the model weights is highly compact, with a size of just 20MB.
Even without any optimization, our implemented prototype character control system runs smoothly at over 60 frames per second.
During runtime, 
when the user changes the style label, the classifier-free guidance on past motion (CFG-PM) is triggered, and we set the guidance scale $\gamma$ to 0.7. \revise{ When the style label remains unchanged, we experimented with various $\gamma$ and found values between 1 and 2 yield similar best results.
Therefore, we set $\gamma$ to 1, which is equivalent to using only the fully conditioned model $\net(\sample_t, t; \pastmotion, \stylelabel, \rootposproj, \rootrotproj)$ to minimize the computational cost while preserving the highest quality. 
We apply $\nbapplied = 15$ generated postures to the character when the user-supplied signals remain unchanged.
}
For the heuristic future trajectory extension, we recycle \revise{$K = 15$} samples of the model-predicted future trajectory. 

\section{Experiments}
In this section, we evaluate the effectiveness of our method for real-time multi-style character control, compare it to other real-time character controllers, and validate the important design choices through ablation studies.
video for qualitative evaluation.

\subsection{Evaluation Setting}
\paragraph{Datasets.}
We conduct experiments on 100STYLE dataset ~\cite{style100}, a publicly available mocap dataset featuring a diverse set of locomotion skills.
The data comprises over 4 million frames of motion capture data encompassing 100 different styles. Our experiments involved training the model using all styles and subsequently testing it within a real-time character control system, where varying control signals are provided.
%

\paragraph{Metrics.}
%
To evaluate our method, we assess the quality and diversity of the motion it generates, the alignment of its output with user-supplied control signals, and its responsiveness to changes in those inputs.
While no single metric can perfectly evaluate the generation, we utilize a combination of the following metrics that are well-established and commonly used in character animation.

To evaluate the motion quality, we use
(1) \emph{Fréchet Inception Distance (FID)}, which is a widely used metric in generative models. It measures the distance between the distribution of the generated motion and the distribution of the training data, offering insights into the fidelity and similarity between the generated and real motion data.
\revise{In our evaluation, we calculate it for each style and then report the average value};
(2) \emph{Foot Sliding Distance}, which is crucial for evaluating the realism and plausibility of the generated motion. It quantifies the moving distance (in meters) of the character's toes when the height of the joint is below a certain threshold (e.g., 1cm).
%
(3) \emph{Acceleration}, the mean per-joint acceleration (in centimeter per second) is calculated to assess the smoothness of the generated motion;
(4) \emph{Diversity (intra-style)}: Given the motions generated with the same control signals (the style and trajectory), this metric computes
the variance of each joint's spatial locations over time, and the average over all joints.

To evaluate the alignment between the generated motion and the control signals, we use 
(5) \emph{Trajectory Error}, which is measured as the angular difference (in degrees) between the expected root moving direction for the subsequent frame and the actual root moving direction of that frame.
(6) \emph{Orientation Error}, which is the difference (in degrees) between the expected and actual generated orientations of the root;
\revise{(7) \emph{Style Accuracy}, each set of generation is divided into 60 frames clips, and we report the percentage of these clips that are consistent with the input style label. }

%
Last, to evaluate the performance of each model in transitioning between different styles, we use
(8) \emph{Transition Duration}, which records the number of frames needed for successful transitions;
(9) \emph{Transition Success Rate}, for which we ask human users to determine if the transition is successfully made within a time window (4s). We report the percentage of successful transitions to a target style.

\paragraph{Baseline methods.}
We compare our method with several state-of-the-art DL-based character controllers.
Motion matching is the de facto standard character controller in the industry. 
The default algorithm utilizes pose and velocity features to search the best match from the database; we enhance the algorithm by incorporating the DeepPhase features introduced in~\cite{deepphase}, denoted as \emph{MM-DP}. 
We assign a weight of 1 to the phase vector to maintain a balanced influence in our feature set.
Due to the high memory-consumption nature of motion matching,
we can only apply it to 10 styles to avoid memory overflow during the construction of KD trees. 
Then we compare against
Local Motion Phase (\emph{LMP})~\cite{localphase} and a variant of combining the Mixture of Experts in MANN~\cite{mann} with DeepPhase features~\cite{deepphase} (denoted as \emph{MANN-DP}), which have shown state-of-the-art performance in recent literature~\cite{deepphase}.
\revise{
Furthermore,
we compare our method against another generative method \--- \emph{MoGlow}~\cite{moglow}. 
Similar to training our model, we apply this method to the entire 100STYLE dataset, and then deploy them to produce motions by concatenating the style label as an extra dimension with others control signals for direct comparisons.
%
Note that the results of MoGlow are generated offline as it does not support real-time character control.
}

\begin{table}[t!]
\centering
\small
\caption{
\revise{Quantitative results on single-style character control. 
The orientation error is not applicable for MoGlow as it does not support facing direction control.}
}

\resizebox{\columnwidth}{!}{
\begin{tabular}{l|ccc|ccc}
\toprule
             & \multicolumn{3}{c|}{Motion Quality} & \multicolumn{3}{c}{Condition Alignment}  \\ \cline{2-7} 
             & FID $\downarrow$  & Ft. Slid. $\downarrow$  & Accel. $\downarrow$  &  Traj. Err. $\downarrow$    & Orient. Err. $\downarrow$   &  Styl. Acc.$\uparrow$   \\ \hline
MM-DP &   3.455  &  1.063  &  1.602  &  53.160  & 19.321  &     23.6\%   \\ 
LMP &  1.255  &  0.872 &  1.098  &  71.691 & \textbf{2.671}  &    51.5\%    \\
MANN-DP & 1.629 & 0.764 &  1.097  & 59.005  & 3.634  &   36.5\%   \\ 
MoGlow &  3.053 & 1.317 &  \textbf{1.009}  & 51.356  &  N/A & 5.2\% \\ \hline
Ours & \textbf{0.913}  & \textbf{0.685} &  1.077  &  \textbf{22.818} & 3.997  &  \textbf{89.5\%}   \\        \bottomrule
\end{tabular}
}
\label{tab:path}
\end{table}

\subsection{Results}
We evaluate the performance of each method with two different settings, namely single-style character control and multi-style character control.
%
In the single-style character control test,
for each style,
we preset a sequence of varying control trajectories, input it to all the competing models with the same style label, and record the rolled-out motions multiple times for evaluating the generated motion in terms of the \emph{quality}, \emph{intra-style diversity}, and \emph{control alignment}.
%
%
To evaluate the ability to deliver inter-style diversity, we conduct experiments using a multi-style control setting. In this setting, we pre-determined a path for the character to follow, updated the style label in the middle, and recorded the transition motions between different styles for evaluation. Specifically, we used each of the 100 styles as the starting style and randomly sampled five styles as the target style for transitioning.


\paragraph{Qualitative Evaluation.}

%
Snapshots of the single-style character control experiments are shown in Figure~\ref{fig:drunk};
the trajectories of the joints produced in real-time using our method and the other baselines based on the same user inputs are presented.
%
%
%
As illustrated in Figure~\ref{fig:drunk}, our character produces much more diversified motions yet consistent with the control inputs.
Baseline methods tend to be trapped in periodic states, resulting in repetitive motions and larger deviations from the control trajectory. 
\revise{
While MoGlow seems to produce smooth movements, it often results in numerous frames sticking, which hampers the generation of high-quality movements(e.g., severe foot sliding during side walking). 
In addition, MoGlow is unable to arbitrarily switch between different styles.
}
%

In the multi-style control transition experiments,
our method can produce natural and agile motions for transitioning between different styles,
even when the transition motions between styles do not exist in the dataset.
In particular,
the neural models have difficulties in switching to a target style that has mirrored motions of those of the original style, as their motion features are symmetric and there is absolutely no transition motions in the dataset.
In Figure~\ref{fig:hop}, 
we select a pair of such styles, namely \quotes{LeftHop} and \quotes{RightHop}, for visual comparisons of all controllers. 
It is evident that the baseline methods fail to switch between these styles, whereas our method successfully completed the switch. 
This can be attributed to our proposed sampling strategy, where we apply CFG on the past motion, which facilitates rapid and seamless style switching.
%


%

More qualitative results can be found in the supplementary video.

\paragraph{Quantitative Evaluation.}

\begin{table}[t!]
\centering
\scriptsize
\caption{Quantitative results of multi-style character control. 
The target style label is given in the middle, and we record the character's motion over the subsequent 120 frames.
}


\resizebox{\columnwidth}{!}{
\begin{tabular}{l|ccc|ccc}
\toprule
             & \multicolumn{3}{c|}{Motion Quality} & \multicolumn{2}{c}{Transition} \\ \cline{2-6} 
             & FID ↓  & Ft. Slid. ↓  & Accel. $\downarrow$   &  Trans. Dura. ↓    &  Succ. Rate ↑     \\ \hline
MM-DP &  3.812 & 1.129  &  1.416 &    69.34 &  56.0\%      \\ 
LMP &  4.193  & 0.418 &  1.055  &    72.22   &    75.6\%     \\
MANN-DP & 2.826 & 0.629 & 1.074  &    82.86 &  62.4\%    \\ 
MoGlow &  2.955   & 1.100 & \textbf{1.017}  &  98.49  &    19.8\%  \\ \hline
Ours &  \textbf{2.281}  & \textbf{0.627}  &  1.049  &   \textbf{48.47}  & \textbf{94.2\%}      \\        \bottomrule
\end{tabular}
}

\label{tab:transition}
\end{table}

We present the quantitative results of single-style character control experiments in Table~\ref{tab:path}, and multi-style transition experiments in Table~\ref{tab:transition}.
From Table~\ref{tab:path}, we can see that
our method outperforms all baseline methods in terms of the motion quality and condition alignment.
We observed that
while motion matching has fewer artifacts, it occasionally fails to find good matches that respond perfectly to the given control signal, causing obvious misalignments.
LMP and MANN share the same autoregressive architecture, allowing them to follow the specified trajectory more 
closely. 
However, their input features are concatenated to form a single embedding, the balance between various signals might be biased under different situations, leading to unstable control, e.g., the style label lost control and the style of the motion is dominated by that of the past motion. 
More specifically, we observed that some styles, such as \quotes{LeftHop}, \quotes{RightHop}, etc., are challenging to transition to from different styles.
Additionally, the generated motion
\begin{wrapfigure}{rh!}{0.3\columnwidth} 
    \centering
    \footnotesize
    \begin{tabular}{l|c}
    \toprule
    & Div. $\uparrow$       \\ \hline
    MM-DP & 0.414   \\ 
    LMP &   0.426   \\
    MANN-DP & 0.353   \\ \hline
    Ours &   \textbf{0.486}     \\        \bottomrule
    \end{tabular}
\end{wrapfigure}
sometimes loses its distinct characteristics, leading to lower FID scores. 
\revise{
The competing generative method, MoGlow, struggles to produce high-quality motions and follow the style specification correctly,
as manifested by worse FID and style accuracy scores.
%
%
}
%
Note the intra-style diversity score is calculated only based on the results obtained from the "waving" and "drunk" styles, as they display a high degree of diversity, 
unlike other styles.
The numerical results are presented in the inset table, where our method has the best diversity.

For multi-style transition experiments, Table~\ref{tab:transition} shows that our method reliably and efficiently transitions to new target styles with fewer transitional frames and a high success rate. 
In contrast to our approach, other baselines frequently struggle to transition to the target style or require many more frames for successful transitions.



\subsection{Ablation Study} 
To show the importance of various algorithmic designs, we compare our full model to several  variants:
\begin{itemize}
  \item \emph{Ours w/o SCT}, which removes the separate condition tokenization and instead learns to map all conditions into a single token as done in~\cite{mdm};
  \item \emph{Ours w/ CFG-S}, which uses the CFG on the style label
  for generating transition motions between different styles;
  \item \emph{Ours w/o HFTE}, which removes the heuristic future trajectory extension module;
  \revise{\item \emph{Ours w/ MLP}, which uses a multilayer perception (MLP) network (8 layers) instead of the transformer.}
\end{itemize}


\begin{table}[t!]
  \centering
  \footnotesize
  \caption{Ablation study.
  Our full model performs best over all ablated versions.
}
    \begin{tabular}{l|ccc|cc}
    \toprule
          & \multicolumn{3}{c|}{Motion Quality} & \multicolumn{2}{c}{Condition Alignment} \\
\cmidrule{2-6}          & FID $\downarrow$   & Ft. Slid. $\downarrow$ & Accel. $\downarrow$ & Traj. Err. $\downarrow$ & Orient. Err. $\downarrow$ \\
    \midrule
    Ours w/o SCT & 1.245 & 0.846 &  1.049 & 31.526 & 5.429 \\
    Ours w/o HFTE & 1.292 & 0.732 & 1.371 & 28.041 & 4.733 \\
    Ours w/ MLP & 1.223 &  0.923 & 1.097  & 51.248  & 7.656 \\
    Ours  & \textbf{0.913} & \textbf{0.685} & \textbf{1.077} & \textbf{22.818} & \textbf{3.997} \\
    \bottomrule
    \end{tabular}%
  \label{tab:token_ablation}%
\end{table}%

Table~\ref{tab:token_ablation} shows that our full model performs best over the ablated versions.
More specifically, 
removing the separate condition tokenization (Ours w/o SCT) and instead representing various control signals as a single toke leads to degraded motion quality (overly smooth motions in particular) and control alignment.
Disabling the heuristic future trajectory extension (Ours w/o HFTE) results in sudden jitters in runtime, which is reflected in the significantly increased Acceleration score.
\revise{Using an MLP as the network backbone (Ours w/ MLP) results in a deterioration of network performance, as evidenced by a lower FID and higher sliding error. }

%
\revise{CFG on the style label (CFG-S), is a straightforward way of generating transitional motions between two different styles. Similar to our CFG-PM, CFG-S is implemented by randomly setting the style vector to nil while keeping all other conditions during training, so we can apply CFG technique on the style label during runtime. 
However, }Table \ref{tab:transition_ablation} shows that applying CFG on the style label leads to failure in transitioning between different styles, whereas our method can achieve remarkable results in quickly and accurately transitioning between different styles of motions, even with the absence of transitional data between distinct styles. 
%
We speculate that during the training phase, past motions contain
rich stylized features that could influence the autoregressive generation of the future motion and prevent it from switching to a new style in the absence of transitional data.
Therefore, by applying the CFG technique to the past motion with a scale factor, we can 
guide the model to rely less on the past motion and instead focus more on the style label, leading to successful transitions.

\begin{table}[b!]
\centering
\scriptsize
\caption{\revise{Ours (w/ CFG-PM) vs. Ours w/ CFG-S.}
Ours w/ CFG-S leads to significantly degraded performance.
}

\resizebox{0.7\columnwidth}{!}{
\begin{tabular}{l|cc}
\toprule
              & \multicolumn{2}{c}{Transition} \\ \cline{2-3} 
               & Trans. Dura.  $\downarrow$   &  Succ. Rate $\uparrow$    \\ \hline
Ours w/ CFG-S   &  110.44 &    8.9\%  \\    \hline
Ours            &  \textbf{48.47}   &  \textbf{94.2\%}      \\        \bottomrule
\end{tabular}
}

\label{tab:transition_ablation}
\end{table}

\revise{
\paragraph{Effect of different diffusion steps.}
In Table \ref{tab:timesteps}, We demonstrate the impact of using different diffusion steps during training and sampling.
It indicates that the 8-step diffusion model is an ideal choice for our real-time character control task, striking a good balance between quality and execution time.
Indeed, this design choice was shaped by the distillation technique inherent in diffusion models.
However, during our initial development, we discovered that training a 1000-step CAMDM first and then distilling it into a fewer-step model led to highly unstable results, whereas directly training fewer-step diffusion models yielded significantly better results. This empirical observation aligns with the finding of MDM~\cite{mdm} as demonstrated in Table 6 in the supplementary and that of AMDM~\cite{amdm}, where a 1000-step diffusion model does not consistently produce optimal results.}

We also conduct experiments to show the effect of changing the number of frames applied to the character during runtime on the motion diversity (see the supplementary video).
\section{Conclusion}

We have presented a novel character controller
that enables real-time generation of high-quality, diverse character animations based on user interactive control, supporting multiple styles with a single unified model. 
This is achieved by a transformer-based
Conditional Autoregressive Motion Diffusion Model, which takes as input
the past motion and can generate a range of diverse potential
future motions conditioned on high-level, coarse user control.
Despite the success demonstrated,
we note a few shortcomings of our method.

\begin{table}[t!]
\centering
\scriptsize
\caption{
\revise{The effect of different diffusion steps.}
}
\resizebox{0.8\columnwidth}{!}{
\begin{tabular}{l|ccccc}
\toprule
Steps        & 2 &  4  &  8 &  16  &  32    \\ \hline
Inf. Time (ms) & 6 &  8 &  13     &  25 &  47     \\
FID↓ & 0.938 & 0.926 & \textbf{0.913}  &   0.919   &  0.914 \\  
Ft. Slid.↓    & 0.832  & 0.780 & 0.685   & \textbf{0.675} &   0.692  \\ 
\bottomrule
\end{tabular}
}
\label{tab:timesteps}
\vspace{-0.3cm}
\end{table} 

While our method can animate characters in real-time using an RTX 3060 GPU, 
it is worth noting that the 8-step diffusion model used in our method is still inherently computationally demanding. 
This presents a significant challenge for its potential adoption on mobile devices, due to their limited computational capabilities.
Hence, it would be worth exploring a more advanced way of accelerating diffusion generation, such as the consistency model~\cite{cm} and latent consistency model~\cite{lcm} that are commonly adopted in image diffusion models.
Moreover, we have demonstrated on mocap data containing only human interactions with flat ground.
It would be intriguing to explore whether the proposed model can effectively process and understand more complex sensory data from the surrounding environment, and then generate natural, adaptive motions in real time that are responsive to changes in the scene.
Lastly, the joystick-based controller, due to its limited expressiveness, may struggle to accurately represent more complex desired motions.
Looking ahead, we aim to explore the potential of developing new control interfaces that integrate signals from various modalities. These could include text prompts, audio instructions, and brain electroencephalogram signals, among others, to enhance the system's ability to generate intricate movements.

\begin{acks}
Taku Komura and Mingyi Shi
are partly supported by Technology Commission
(Ref:ITS/319/21FP) and Research Grant Council (Ref:
17210222), Hong Kong.
\end{acks}

\bibliographystyle{ACM-Reference-Format}
\bibliography{bibliography.bib}

\appendix
\newpage

\begin{figure}[h!]
  \centering
  \includegraphics[width=\linewidth]{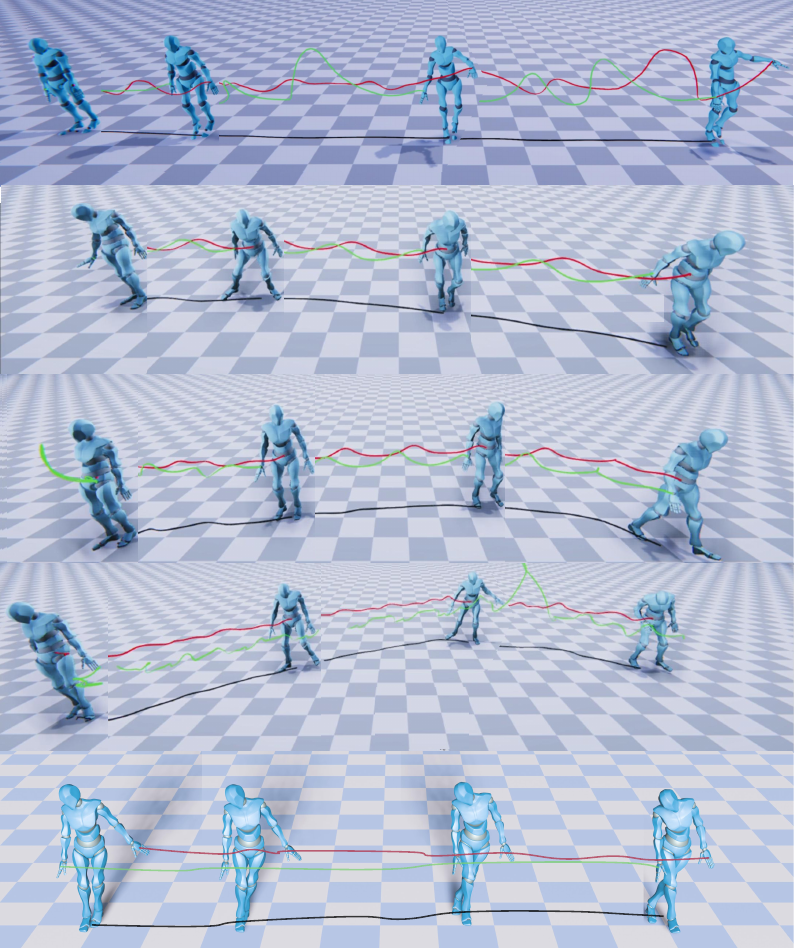}
  \caption{
  Visual comparisons of single-style control. 
  From top to bottom: Ours, LMP, MANN-DP, MM-DP and MoGlow, all using the same control inputs. The trajectory of the left hand, right hand, and root are colored in red, green, and black, respectively. Observe the high diversity and better adherence to the control trajectory of our method compared to the other baselines.
  }
  \label{fig:drunk}
\end{figure}

\begin{figure}[h!]
  \centering
  \includegraphics[width=\linewidth]{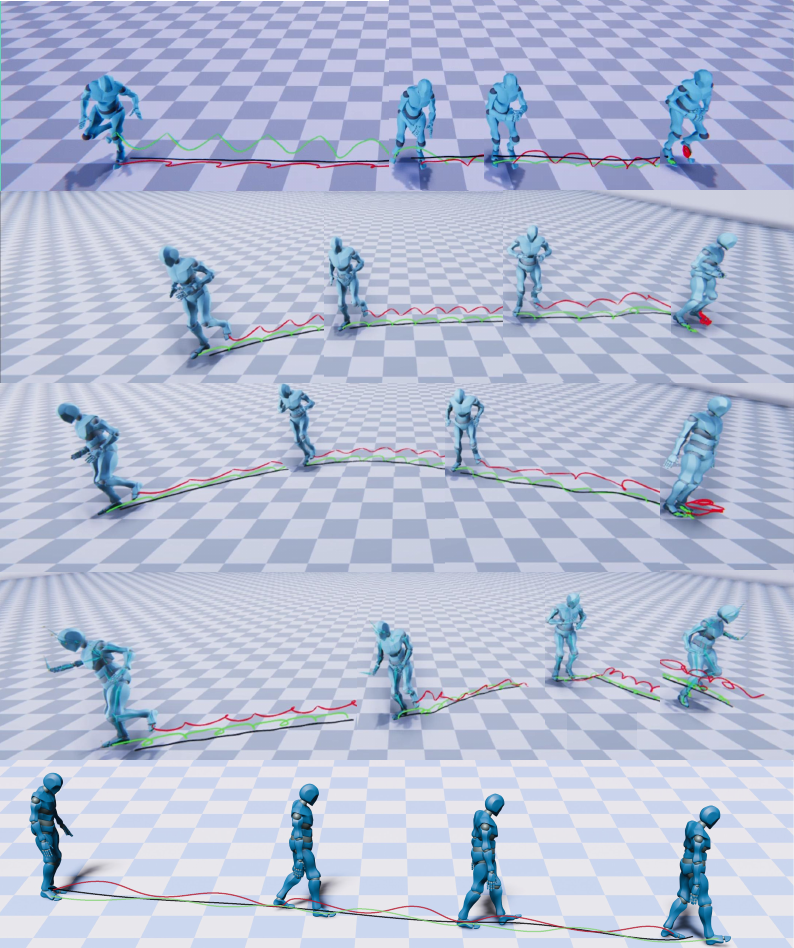}
  \caption{
  Visual comparisons of multi-style control.
  From top to bottom: Ours, LMP, MANN-DP, MM-DP and MoGlow.
  Our method can transition naturally between distinct styles (\quotes{LeftHop} and \quotes{RightHop} in this case), whereas baselines fail.
  }
  \label{fig:hop}
\end{figure}

\newpage




\end{document}


\title{Supplementary material: \\
Example-based Motion Synthesis via Generative Motion Matching
}

\author{First Author}
\email{firstauthor@example.com}
\orcid{1234-5678-9012}
\affiliation{%
  \institution{First Institution}
  \country{Firstcountry}
}

\author{Second Author}
\orcid{1234-5678-9012}
\affiliation{%
  \institution{Second Institution}
  \country{Secondcountry}
}

\renewcommand{\shortauthors}{Firstauthor et al.}
\newcommand{\name}{GenMM\xspace}

\definecolor{red}{rgb}{0.8, 0.2, 0.2}
\definecolor{purple}{rgb}{0.99,0.2,0.72}

\newif\ifdraft
\drafttrue

\ifdraft
\newcommand{\xl}[1]{{\color{blue}[xl: #1]}}
\newcommand{\pl}[1]{{\color{red}{#1}}} 
\newcommand{\plc}[1]{{\color{red}{[Peizhuo: #1]}}} 
\newcommand{\wy}[1]{{\color{orange}[Weiyu: #1]}}
\newcommand{\osh}[1]{{\color{purple}{#1}}} 
\newcommand{\OSH}[1]{{\color{purple}[Olga: #1]}}
\newcommand{\bq}[1]{{\color{green}[bq: #1]}}
\newcommand{\todo}[1]{{\color{yellow}[todo: #1]}}

\else
\newcommand{\xl}[1]{}
\newcommand{\pl}[1]{{\color{black}{#1}}}
\newcommand{\osh}[1]{{\color{black}{#1}}} 
\newcommand{\plc}[1]{} 
\newcommand{\wy}[1]{}
\newcommand{\OSH}[1]{}
\newcommand{\bq}[1]{{\color{green}[bq: #1]}}
\newcommand{\todo}[1]{{\color{yellow}[todo: #1]}}

\fi

\maketitle

\newpage



